\documentclass[preprint,aps,epsfig,floats,superscriptaddress,nofootinbib]{revtex4-1}
\usepackage{pstricks,pstricks-add}
\usepackage{amsfonts}
\usepackage{color}
\usepackage{mathtools}
\usepackage{braket}
\usepackage{graphicx,epsf,natbib,amsmath,latexsym,amssymb,multirow}
\usepackage{epstopdf}
\usepackage{simplewick}
\usepackage{hyperref}
\newcommand{\be}{\begin{equation}}
\newcommand{\ee}{\end{equation}}
\newcommand{\bea}{\begin{eqnarray}}
\newcommand{\eea}{\end{eqnarray}}
\newcommand{\ba}{\begin{array}}
\newcommand{\ea}{\end{array}}

\def \nn {\nonumber}
\newcommand{\eq}[1]{(\ref{#1})}

\newcommand{\s}{\sigma}

\newcommand{\beq}{\begin{eqnarray}}
\newcommand{\eeq}{\end{eqnarray}}
\newcommand{\bes}{\begin{subequations}}
\newcommand{\ees}{\end{subequations}}

\begin{document}

\title{Towards Holographic Quantum Energy Teleportation}

\author{Dimitrios Giataganas}\email{dgiataganas@phys.cts.nthu.edu.tw}
\affiliation{Physics Division, National Center for Theoretical Sciences,
National Tsing-Hua University, Hsinchu, 30013, Taiwan}
\author{Feng-Li Lin}\email{fengli.lin@gmail.com} \author{Pei-Hua Liu}\email{lphhpl@gmail.com}
\affiliation{Department of Physics, National Taiwan Normal University, Taipei, 11677, Taiwan}

\begin{abstract}

   We propose a simplified protocol of quantum energy teleportation (QET) for holographic conformal field theory (CFT) in 3-dimensional anti-de Sitter space with or without black hole.  As a tentative proposal, we simplify  the standard QET by replacing Alice's local measurement with the local projection, which excites the system from ground state into a particular state dual to a Banados geometry. We then mimic Bob's local operation of the usual QET for extracting energy by deforming the UV surface with a local bump. Adopting the surface/state duality this deformation corresponds to local unitary. We evaluate the extraction of energy from the holographic stress tensor, and find that Bob always gains energy extraction in our protocol. This could be related to the positive energy theorem of the dual gravity. Moreover, the ratio of extraction energy to injection one is a universal function of the UV surface deformation profile. 
\end{abstract}

\maketitle
\tableofcontents

\section{Introduction}

Quantum entanglement as powerful resources for communication or computation was firstly demonstrated by the quantum state teleportation protocol \cite{Bennett:1992tv,Buhrman} and for the discussions in the quantum field theory context, see e.g. \cite{Laiho:2000mc}.  This is a bipartite protocol with a EPR/Bell pair, e.g., $|\Phi^+\rangle:={1\over \sqrt{2}} (|00\rangle \pm |11 \rangle)$, shared by Alice and Bob who are distantly separated. To send an unknown quantum state in Alice's hand to Bob without actually sending the qubit, Alice first perform a local joint measurement on her two qubits in the Bell-state basis, i.e., the so-called Bell-state measurement. This measurement will then maximally entangle Alice's two qubits. By the monogamy of quantum entanglement \cite{Coffman:1999jd}, Bob's qubit is then disentangled from Alice's, state of which should however depends on Alice's outcome of her Bell-state measurement. Alice then send her measurement outcome to Bob by classical communication. Based on the received message,  Bob can perform some appropriate local unitary operation on his qubit to recover Alice's unknown state.

From the above description, we can see that the essential ingredients for quantum state teleportation are (i) shared quantum entanglement as resources, (ii) local measurements, (iii) local unitary operations and (iv) classical communications. Note that (iii) and (iv) are the so-called local operations and classical communication (LOCC). In fact, these are also the ingredients for the performing the task of quantum energy teleportation (QET) \cite{Hotta:2008uk,Hotta:2009fz,Hotta:2010py,Hotta:2011xj}, which will be the focus of this work.  Instead of sending quantum information as in quantum state teleportation, it is the energy to be teleported in the QET task.

   The QET protocol is very similar to quantum state teleportation, and in some sense even simpler. We need to have a (many-body) quantum state of nontrivial entanglement. Alice performs local  measurements,  which then excite the state by injecting the energy and build up an energy density profile extending far beyond Alice's local region due to the quantum entanglement. A distant Bob then need to perform some local unitary operations to extract the energy out of the tails of energy density profile of the excited state. As in the quantum state teleportation, in general the LOCC is needed in order for Bob to do the right local operations according to Alice's measurement outcomes to gain the energy.  By the passivity of the quantum states the Bob's energy gain should be less than the amount of Alice's injected energy in order not to violate the second law of thermodynamics \cite{Pusz:1977hb}.

   In this work, we propose a holographic version of simplified QET protocol in the context of AdS$_3$/CFT$_2$ correspondence, and obtain the relation between Bob's energy extraction gain and Alice's injected energy density.  Since there is no known holographic dual realization of the local measurement, we replace Alice's local measurement in the standard QET by local projection. In this protocol the state after Alice's projection is fixed, thus there is no need for classical communication. Bob can perform arbitrary local operation to extract energy.   Even with such kind of simplification, our protocol is shown to be able for Bob to gain the energy extraction. This success shall inspire more studies on the holographic realization of nontrivial quantum information tasks.

    Our paper is organized as follows. In the next section, we will review a toy model of QET protocol, which motivates our proposed protocol. In the  section \ref{sec 3} we  describe in details our holographic QET protocol in AdS$_3$, and in section \ref{sec 4} we generalize this protocol to the finite temperature case in BTZ background \cite{Banados:1992wn}. We then conclude our paper.

\section{Toy QET model}

 In this section we will describe a toy QET model in \cite{Hotta:2010py,Hotta:2011xj} and its variation to motivate and illustrate our setup for the holographic QET.

\subsection{A qubit model for QET}

We now review a qubit model for QET proposed by Hotta in \cite{Hotta:2010py,Hotta:2011xj}  to set up the platform  to motivate our construction of holographic QET protocol. This is a minimal model for QET with Alice and Bob sharing a pair of entangled qubits, which is the ground state of the following Hamiltonian:
\be
H_{T}=H_A + H_B +V
\ee
with
\be
H_A:=h \; \sigma_3 \otimes I_{2\times 2} + {h^2 \over \sqrt{h^2+k^2}}\; I_{4\times 4},\qquad H_B:=  h\; I_{2\times 2} \otimes   \sigma_3+ {h^2 \over \sqrt{h^2+k^2}}\; I_{4\times 4}
\ee
and
\be
V:= 2k \; \sigma_1 \otimes \sigma_1 + {2 k^2 \over \sqrt{h^2+k^2}}\; I_{4\times 4}\;,
\ee
where $\sigma_i$'s are the Pauli's matrices, and $I_{2\times 2}$ and $I_{4\times 4}$ are the identity matrices, respectively.

We will assume $h,k \ge 0$, and the constant shifts in the above are just for convenience  such that
\be
\langle g| H_A |g \rangle = \langle g| H_B |g \rangle = \langle g| V |g \rangle =0
\ee
where $|g\rangle$ is the entangled ground state solved from $H_T|g\rangle =0$.

Alice then performs the local projection measurement by acting on $|g\rangle$ with the projection operator $P_A[\alpha]:={1\over 2}( I_{4\times 4} + \alpha\; \sigma_1 \otimes I_{2\times 2})$ with $\alpha=\pm 1$.  The post-measurement state with outcome $\alpha$ is $|M(\alpha)\rangle:={1\over \sqrt{p_A[\alpha]} }P_A[\alpha]|g\rangle$ where  $p_A[\alpha]:=\langle g| P_A[\alpha] |g\rangle$ is the probability for the outcome to be $\alpha$. In the current case, $p_A[\alpha]=1/2$.  The projection measurement with outcome $\alpha$ will then inject to the system the following amount of energy
\be
E_A[\alpha]=\langle g| P_A[\alpha] H_T P_A[\alpha] |g\rangle= {h^2 \over 2 \sqrt{h^2+ k^2}}\;.
\ee
 We see that $E_A[\alpha]$ is non-negative as expected. Then, the injected amount of energy $\Delta E_A$ is just $\sum_{\alpha} p_A[\alpha] E_A[\alpha]=E_A[\pm 1]$. \footnote{Note that  in the above $E_A[\alpha]$ is  independent of $\alpha$ due to the particular choice of the Hamiltonian and the projection operator. Generically, the amount of energy change depends on the measurement outcome if some other projection operators are chosen.}

Next step of QET protocol is for Bob to perform some local operation on his qubit state just according to Alice's measurement outcome. In this case, one chooses the local operation to be
\be
U[\theta; \beta]:=e^{-i \; \beta \; \theta \; I_{2\times 2}\otimes \sigma_2}
\ee
where $\beta$ can take the values $\pm 1$ and is not fixed yet. The other  parameter $\theta$ is real, measuring the amount of rotation.   Note that this local operation will not change Alice's qubit state as the post-measurement state $|M(\alpha)\rangle$ is not entangled. Thus, energy extraction is completely from Bob's local qubit. Then, given a set of parameters $\theta$, $\alpha$ and $\beta$, the amount of energy extraction is
\bea\nn
E_B[\theta; \alpha, \beta]&:=&E_A[\alpha] - \langle g| U^{\dagger}[\theta; \beta] P_A[\alpha] H_T P_A[\alpha] U[\theta; \beta] |g\rangle \\ \label{EBab}
&=& { \sin\theta [ \alpha \beta h k \cos\theta - (h^2 + 2 k^2) \sin\theta ] \over \sqrt{h^2+k^2}}\;.
\eea

\begin{figure}
\includegraphics[width=.48\columnwidth]{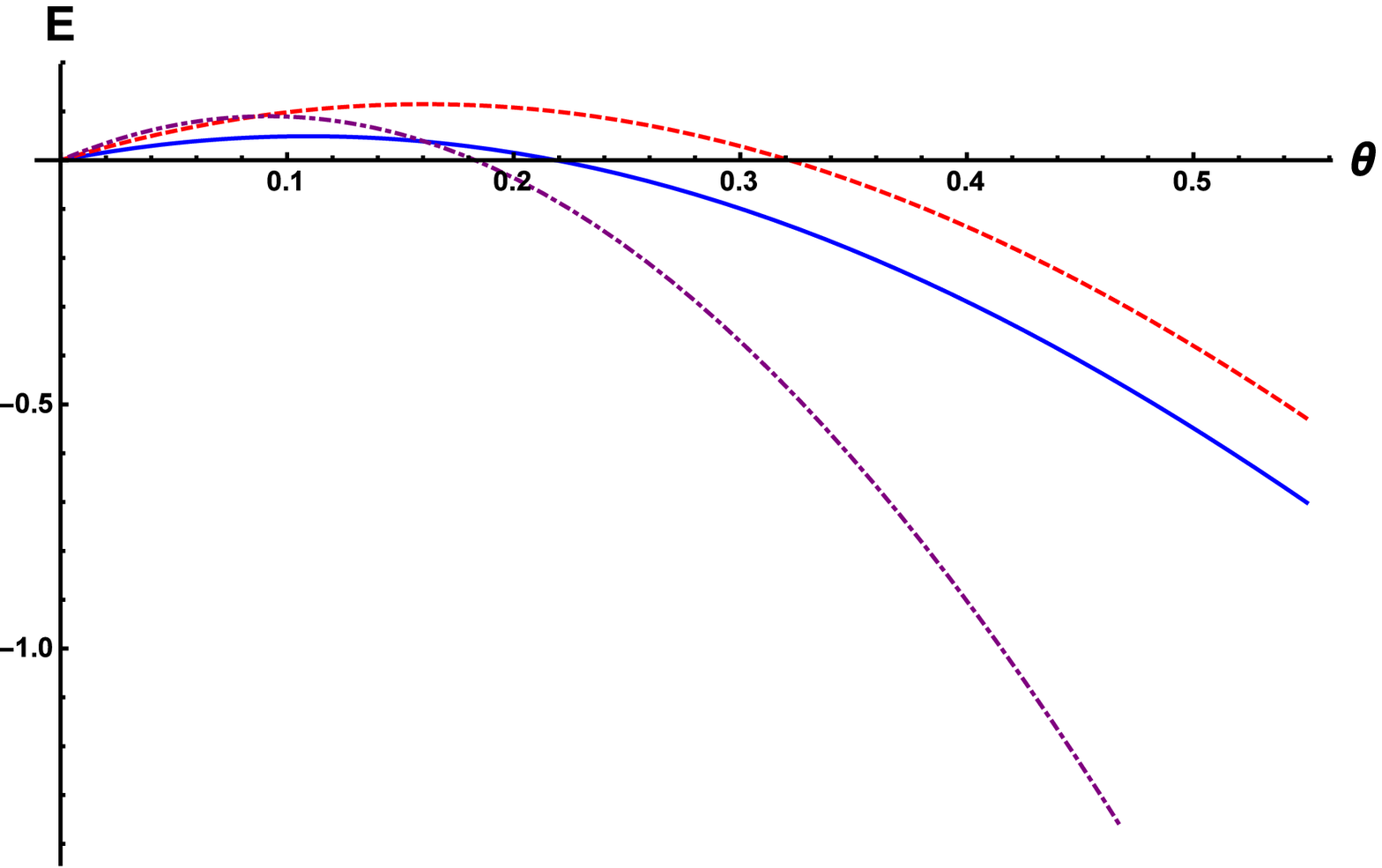}\hspace{.5cm}
\includegraphics[width=.48\columnwidth]{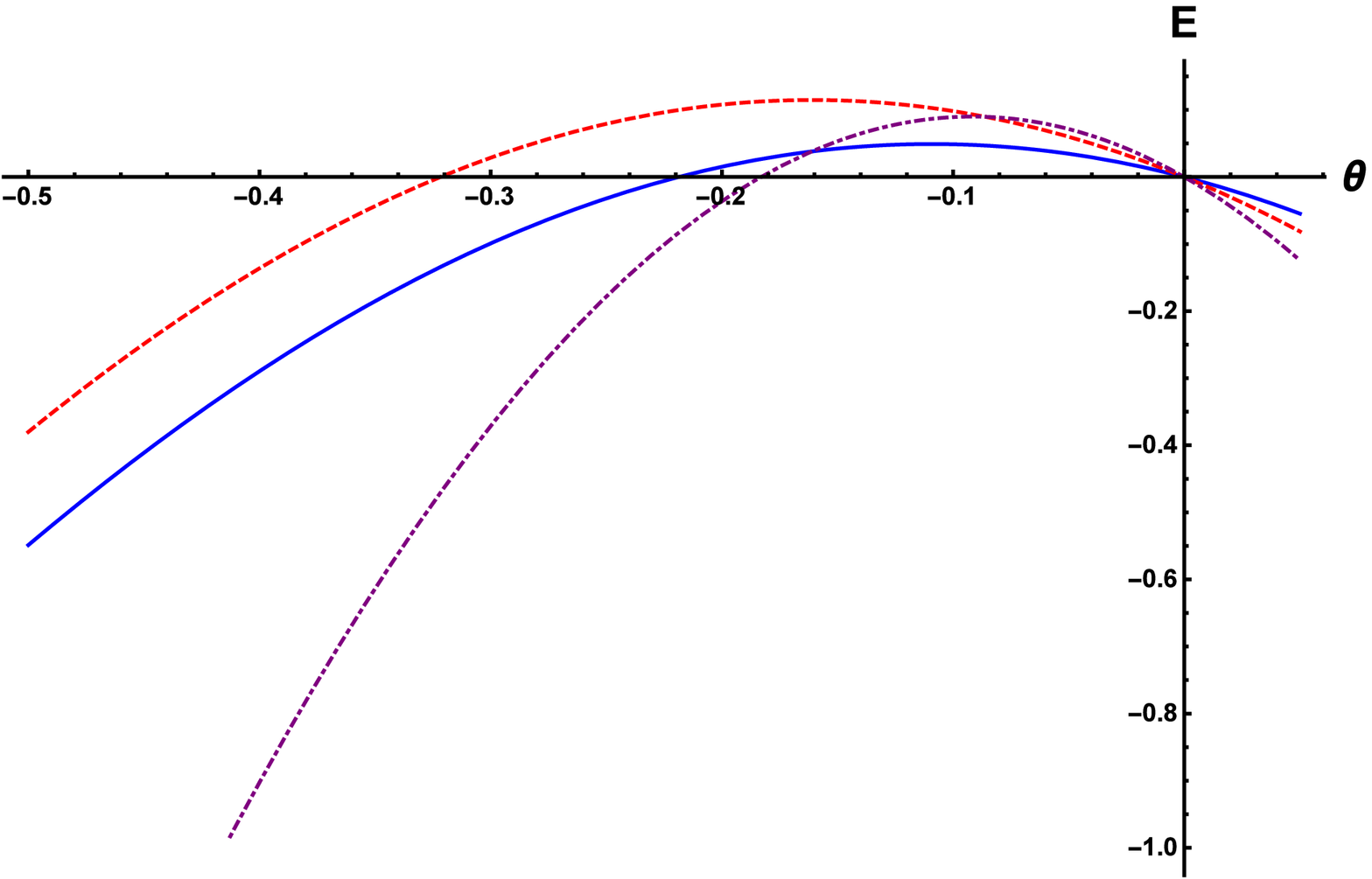}
\caption{Energy extraction of qubit model for QET with respect to the tuning of local unitary operation, i.e., Eq. \eq{EBab}: (Left panel) $\Delta E_B|_{\alpha\beta=1}$ v.s. $\theta$, and (Right panel) $\Delta E_B|_{\alpha\beta=-1}$ v.s. $\theta$. We set $k=1$ and $h=0.5$ (blue-solid), $1$ (red-dashed), $5$ (purple-dotdashed). In all cases there is a window for Bob to gain the energy extraction.}
\label{fig:ener1}
\end{figure}

From \eq{EBab} we can obtain the average energy extraction for the QET protocol with $\alpha\beta=\pm 1$ \cite{Hotta:2010py,Hotta:2011xj}:
\be\label{DEB-t}
\Delta E_B|_{\alpha\beta=\pm 1}:=p_A[+1] E_B[\theta; +1, \pm 1]+p_A[-1] E_B[\theta; -1, \mp 1]=E_B[\theta; +1,\pm 1]\;.
\ee
In Fig. \ref{fig:ener1} we plot the above results as functions of $\theta$ for some given $(h,k)$ and show that there is a window for which $\Delta E_B$ is positive.

   In contrast, if we do not feedback Alice's measurement outcomes to guide Bob's local operations, i.e., $\alpha$ and $\beta$ are not correlated, we will have the amount of extraction energy
\be
 \Delta E_B|_{\beta=1} := p_A[+1] E_B[\theta; +1, +1]+p_A[-1] E_B[\theta; -1, +1] = -{(h^2+2k^2)\sin^2\theta \over \sqrt{h^2+k^2}}\le 0\;.
\ee
As $ \Delta E_B|_{\beta=1}$ cannot be positive, this means that one cannot extraction energy out of LOCC if there is no quantum feedback control.

   In the above discussions, we see that the feedback to correlate Alice's outcome and Bob's local operation is crucial for the QET to succeed. However, there is no known realization of local measurements in the holographic setup, thus it is hard to completely construct the full QET holographically. To proceed we shall simplify the protocol.  One hint for simplification is to note that in \eq{DEB-t} the average extraction energy is the same as  the extraction energy for the projected state with $\alpha=1$. Although this coincidence is due to our particular choice of the Hamiltonian and projection operator, it may suggest that we can simplify the QET by replacing Alice's local measurement by local projection. In some sense, we just consider the extraction of energy out of a particular excited state, and strictly speaking it is not a protocol because Alice and Bob are not instructed to perform particular operations. However, we will still loosely use the word ``protocol".  For such protocol there is no need for classical communication and Bob can perform arbitrary operation to extract energy out of the projected state. A priori there is no guarantee that this simplifying protocol will work, but we will adopt it to construct the holographic realization yielding positive outcome.

\subsection{Schematic holographic QET protocol}

  Guided by the previous discussions, we now attempt to construct the holographic toy model for QET. The difficulty of constructing the holographic QET is two folds.
The first level of difficulty is that the holographic model is a field theory one, in contrast to the qubit model discussed in the above.  Despite that, the QET protocols for the free quantum field theory based on the general theory of quantum measurement have been considered in \cite{Hotta:2008uk,Hotta:2009fz,Hotta:2011xj,Hotta:2009ppa}.  However, it is still hard to construct positive-operator valued measure (POVM) and the local unitary operations (LUs) in the strongly coupled holographic dual CFTs.  The POVM consists of a set of hermitian positive semidefinite operators, not necessarily to be the projectors, which sum to the unity with respect to some chosen Hilbert space, i.e., the completeness. Unlike the POVM for the qubit systems of finite dimensional Hilbert space for the usual quantum information task,  the one for CFT  is with respect to infinite dimensional Hilbert space. Naively, one can use the primary operators as the set of POVM. However, the measure is singular as the set is defined at one point. This can be rectified by smearing over a finite region but it is not clear what smearing is more natural and proper. Another difficulty is not clear how to define the completeness of the chosen POVM in the CFT context for a given quantum information task.    Moreover, it is also not clear how to define a measurement operation which can yield different possible outcome in the sense of some field-theoretical ``qubit".  See \cite{Numasawa:2016emc,Rajabpour:2015uqa} for some pioneer discussions along this direction.

  %dg7
%  The POVM consists of a set of hermitian positive semidefinite operators, other than projectors, on a Hilbert space who sum to the unity. In principle can be thought as acting to system states associated to probe states, producing the final measured state. In the context of holography or CFTs, is not straightforward to realize this setup. A possible candidate for constructing such a POVM, may be with the use of the OPE blocks.

 The second level of difficulty is that we should find the geometrical realizations of the POVM and LUs so that we can calculate the local extraction energy from the bulk gravity side. As the explicit forms of both the POVM and LUs depend on the microscopic details of the CFT, it is quite challenging to construct their holographic counterparts. In the AdS/CFT correspondence, the holographic quantities are usually the coarse-graining of their CFT counterparts. For example, the Ryu-Takayanagi formula \cite{Ryu:2006bv,Ryu:2006ef} gives the entanglement entropy but not its fine-graining, i.e., the entanglement spectrum or reduced density operator.

   To by pass these difficulties, we will adopt the following strategy to construct the holographic QET protocol.  The pictorial representation of our proposed holographic QET protocol is summarized in Fig. \ref{fig:hologr}.

\psset{unit =0.75cm, linewidth=1.5\pslinewidth}
\begin{figure}[!ht]
    \centering
    \begin{pspicture}(-2,-2)(10,10)
    \put(-1,9.5){\fontsize{16pt}{16pt}\bf  Step 1}
\pscurve[linewidth=2pt](0,7.5)(1,9)(2,7.7)(3,8.5)(4,7.7)(5,7.9)(6,7.5)(7,7.7)(8,7.6)(9,7.5)
\psline[linewidth=2pt]{->}(0,7.3)(1,7.3)
\put(1.1,7.2){\bf \Large x }
\psline[linewidth=2pt]{->}(0,7.3)(0,8.3)
\put(-0.6,8.6){\fontsize{12pt}{12pt}\bf $\rho$(x) }
\psline[linewidth=2pt,linestyle=dashed](1,5)(1,9)
\psline[linewidth=2pt,linestyle=dashed](3,9)(3,5)
\put(1,6.3){\fontsize{18pt}{18pt}\bf LPO}
\psline[linecolor=red,linewidth=2pt](0,5)(1,5)
\psline[linecolor=red,linewidth=2pt,linestyle=dashed](1,5)(3,5)
\psline[linecolor=red,linewidth=2pt](3,5)(9,5)
\put(9,4.8){\bf UV }
\psline[linewidth=2pt]{->}(0,4.7)(1,4.7)
\put(1.1,4.5){\bf \Large x }
\psline[linewidth=2pt]{->}(0,4.65)(0,5.7)
\put(-0.1,5.9){\bf\Large z }
\psline[linecolor=green,linewidth=2pt](0.25,6)(1,6)
\psline[linecolor=green,linewidth=2pt,linestyle=dashed](1,6)(3,6)
\psline[linecolor=green,linewidth=2pt](3,6)(9,6)
\psline[linecolor=blue,linewidth=2pt](0.1,5.5)(1,5.5)
\psline[linecolor=blue,linewidth=2pt,linestyle=dashed](1,5.5)(3,5.5)
\psline[linecolor=blue,linewidth=2pt](3,5.5)(9,5.5)

\put(0.9,3.5){\fontsize{16pt}{16pt} \textbf{\textit{Alice}}}
\put(6,3.5){\fontsize{16pt}{16pt} \textbf{\textit{Bob}}}

\put(-1,2.5){\fontsize{16pt}{16pt} \bf Step 2}
\psline[linewidth=2pt,linestyle=dashed](6,-2)(6,2)
\psline[linewidth=2pt,linestyle=dashed](8,2)(8,-2)
\pscurve[linewidth=2pt](0,0.5)(1,2)(2,0.7)(3,1.5)(4,0.7)(5,0.9)(6,0.5)(7,1.2)(8,0.6)(9,0.5)
\psline[linewidth=2pt]{->}(0,0.3)(1,0.3)
\put(1.1,0.2){\bf \Large x }
\psline[linewidth=2pt]{->}(0,0.3)(0,1.3)
\put(-0.6,1.6){\fontsize{12pt}{12pt}\bf $\rho$(x) }
\psline[linecolor=red,linewidth=2pt](0,-2)(1,-2)
\psline[linecolor=red,linewidth=2pt,linestyle=dashed](1,-2)(3,-2)
\psline[linecolor=red,linewidth=2pt](3,-2)(6,-2)
\psline[linecolor=red,linewidth=2pt](8,-2)(9,-2)
\pscurve[linecolor=red,linewidth=2pt](6,-2)(7,-1.5)(8,-2)
\put(9,-2.2){\bf UV }
\psline[linewidth=2pt]{->}(0,-2.3)(1,-2.3)
\put(1.1,-2.5){\bf \Large x }
\psline[linewidth=2pt]{->}(0,-2.35)(0,-1.3)
\put(-0.1,-1.1){\bf\Large z }
\psline[linecolor=green,linewidth=2pt](0.25,-1)(1,-1)
\psline[linecolor=green,linewidth=2pt,linestyle=dashed](1,-1)(3,-1)
\psline[linecolor=green,linewidth=2pt](3,-1)(6,-1)
\psline[linecolor=green,linewidth=2pt](8,-1)(9,-1)
\pscurve[linecolor=green,linewidth=2pt](6,-1)(7,-0.5)(8,-1)
\psline[linecolor=blue,linewidth=2pt](0.1,-1.5)(1,-1.5)
\psline[linecolor=blue,linewidth=2pt,linestyle=dashed](1,-1.5)(3,-1.5)
\psline[linecolor=blue,linewidth=2pt](3,-1.5)(6,-1.5)
\psline[linecolor=blue,linewidth=2pt](8,-1.5)(9,-1.5)
\pscurve[linecolor=blue,linewidth=2pt](6,-1.5)(7,-1)(8,-1.5)
\put(6.3,-0.5){\fontsize{18pt}{18pt}\bf LU}

    \end{pspicture}
    \caption{Pictorial representation of our proposed simplified QET protocol in holography. (1) The first step of the protocol is for Alice to perform a local projection such that a positive energy density profile is injected. (2) The second step is for Bob to perform local operation to extract energy from the injected energy density profile. In our consideration, the excited state due to Alice's local projection is described by Banados geometry, i.e., asymptotic AdS$_3$ space, and Bob's local operation is described by the deformation (the bump in the lower graph) of UV surface in accordance with the interpretation of surface/state duality, i.e., AdS/MERA duality. According to surface/state duality, each bulk surface is associated with a quantum state in the dual CFT, e.g., the three color lines correspond to three different CFT states, the red one is specified as the UV fixed point state and the other twos are the RG-flowed states.}
    \label{fig:hologr}
 \end{figure}
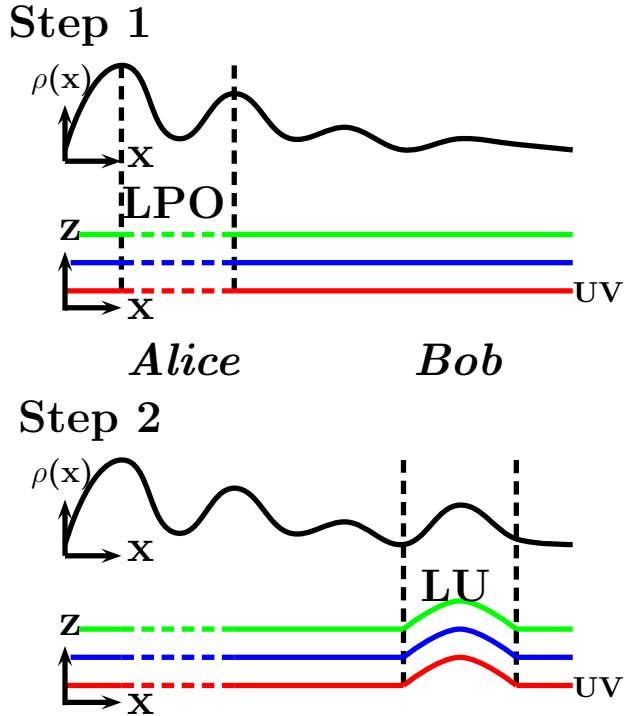

    Firstly,  motived by the discussions in the end of last subsection, we will not consider the projection measurement or POVM. Instead, we just consider the projection of the local degrees of freedom on a finite region $\mathcal{A}$ onto some particular state. After that, the ground state will then be excited with local energy injection around region $\mathcal{A}$.  We will denote this procedure as the ``local projective operation" (LPO) on region $\mathcal{A}$.  As the LPOs could be quite generic, it may not be easy to find the holographic duals.  In this paper, we will focus on a particular class of LPO which can be realized by a conformal map from the (upper-half) complex plane, which has zero stress tensor, to the one with branch cuts in the region $\mathcal{A}$.  The bulk geometry dual to such kind of excited states belong to the Banados geometries \cite{Banados:1998gg}. In particular, we will consider the LPO recently proposed in \cite{Numasawa:2016emc,Rajabpour:2015uqa,Rajabpour:2015xkj} by projecting the local state into Cardy state  in the boundary conformal field theory (BCFTs)  \cite{Cardy:2004hm}.

    We shall emphasize that strictly speaking we are considering extraction of energy out of a particular state, not an ensemble of states as in the standard QET protocol. As the measurement outcome is fixed all the time, there is no need to perform classical communication to correlate Alice's outcome with Bob's local operation. However, we will still loosely consider such a scheme as kind of protocol.  

   Secondly, we need to find the geometrical realization of LUs. At first sight, this seems an even difficult task than constructing the holographic counterpart of LPO as most physical observables are hermitian not unitary. One well-known unitary operator is the time evolution operator which can also be realized by the appropriate conformal map \cite{Calabrese:2005in,Calabrese:2006rx} in 2-dimensional CFT, so is its holographic counterpart, e.g. see \cite{Nozaki:2013wia}. However, for the purpose of QET the LUs are needed to be non-commuting with the Hamiltonian and localized in space. Thus, the time evolution operator is not what we need so that we need to find the other possibilities.
   
   Up to this point, the QET protocol described above should be applied to any 2D CFT, including holographic CFTs because only Virasoro symmetry is used to construct either POVM or LU.  On the other hand, there are hypothetical LUs appearing in the construction of many-body ground state by the so-called multi-scale entanglement renormalization ansatz (MERA) \cite{Vidal:2008zz,Haegeman:2011uy}. In this construction, the microscopic ground state is related to the simpler IR one by a network of LUs. In practice, these LUs are determined by the variational principle. There are two types of LUs in MERA: one is called the disentangler which is to remove the local entanglement by changing basis, and the other is called isometry which is to decouple the unentangled UV degrees of freedom from the IR one, i.e., coarse-graining without truncating the Hilbert space. These LUs are usually localized in space and not commuting with the Hamiltonian, otherwise they will not change the microscopic ground state. Therefore, they will be the candidate LUs for QET protocols. Moreover, later on the MERA network is suggested to have the AdS-like geometry \cite{Swingle:2009bg}, especially by comparing the causal cut of MERA and the Ryu-Takayanagi formula when evaluating the entanglement entropy.

    Based on the AdS/MERA duality, one can further postulate that each co-dimensional two hypersurface in the AdS space is dual to a CFT state, which is related to the UV ground state by MERA's LUs. This is the so-called surface/state duality as proposed in \cite{Miyaji:2015yva,Miyaji:2015fia}, especially it was shown that the Ishibashi boundary states of CFT have no real-space entanglement \cite{Miyaji:2014mca} so that they are proposed to be dual to a point in AdS space. Moreover, in \cite{Huang:2015xca} the disentanglers and isometries can be holographically identified through the integral geometrical decomposition of Ryu-Takayanagi formula.

    After lengthy elaboration on the surface/state duality, we then propose that the geometrical realization of MERA's LUs used in our QET protocol is to deform the UV surface around a local region $\mathcal{B}$ so that it causes a bump around $\mathcal{B}$ into the AdS space. This new surface will then be identified as the new CFT state after Bob performs his LUs. Thus, the QET energy extraction will then be encoded in the extrinsic curvature of the bump, which is related to the change of the boundary stress tensor. Then, we can tune the shape of the bump on the bulk geometry dual to the LPO, and see if there is a chance to obtain positive $\Delta E_{\mathcal B}$.

\section{Towards holographic QET}\label{sec 3}

In this section we would like to implement our tentative holographic QET protocol in AdS$_3$ space.

\subsection{Holographic local projection operation}

    The general measurement used in the QET protocol is to utilize the positive-operator-valued measure (POVM) (see e.g. \cite{Hotta:2011xj}), i.e., to find a set of positive semi-definite operators $\{ F_k \}$ satisfying the completeness condition
\be\label{POVM_1}
\sum_{i=1}^N F_k = \mathbb{I}\;.
\ee
so that for any (mixed) state $\rho_S$ being measured,
\be
\textrm{p}_k=\textrm{tr} (\rho_S F_k)
\ee
is nonnegative and less than one so that it can be interpreted as the probability for the measurement outcome $k$. Moreover, \eq{POVM_1} is then turned into
\be\label{POVM_2}
\sum_k \textrm{p}_k=1\;.
\ee

   In practice, we need to find a set of local projection operators (LPO) $\{ F_k \}$, which act only the local region $\mathcal A$ around Alice. In a discrete lattice system, the LPO can be realized as follows
\be
F_k =\prod_{i\in {\mathcal A}} O_k(i) |0_i\rangle \langle 0_i| O_k^{\dagger}(i)  \otimes \prod_{j\in {\mathcal A}^c} I_j
\ee
where $i$ and $j$ label the lattice sites, $|0_i\rangle$ is the canonical reference state at site $i$, and $O_k(i)$ is some hermitian operator (labelled by $k$) acting on the site $i$.   Note that ${\mathcal A}^c$ is the complement of region $\mathcal A$.

One can then generalize the above to the field theory by taking the continuum limit so that the LPO's in such cases are nonlocal operators.
Moreover, if we consider LPO or POVM in CFTs, then it further requires that these operators are also conformal invariant. Recently there are such kind of operators constructed in CFTs such as conformal blocks or OPE blocks \cite{SimmonsDuffin:2012uy,Czech:2016xec}, we are currently investigating the possibility of utilizing them as POVM for general quantum teleportation tasks \cite{POVM-CFT}.

Instead of worrying about the construction of LPO or POVM, in this work we will focus more on the possibility of realizing the holographic QET protocol with the gain of energy extraction by following our schematic plot in Fig. \ref{fig:hologr}. Therefore, we basically like to know the resultant state after LPO so that we can obtain the amount of the injected energy density profile. Moreover, we will only consider the holographic CFT$_2$.  In the context of AdS$_3$/CFT$_2$  it is known that all the CFT states are dual to the Banados geometries \cite{Banados:1998gg}, i.e.,
\be\label{master metric}
ds^2=R^2 \Big\{\frac{dz^2}{z^2}+\frac{6}{c}  T^{\mathcal{P}}(w)dw^2+\frac{6}{c} \overline{T}^{\mathcal{P}}(\overline{w})d\overline{w}^2+\Big( \frac{1}{z^2}+ z^2 {36\over c^2} T^{\mathcal{P}}(w)\overline{T}^{\mathcal{P}}(\overline{w}) \Big) dw d\overline{w}\Big\}
\ee
where $T^{\mathcal{P}}$ and $\overline{T}^{\mathcal{P}}$ are the stress tensor associated with the excited state of CFT, i.e., in the current context, it is the resultant state after LPO acting on the ground state. Note that the central charge of the CFT is related to the 3D Newton constant $G_N$ and the AdS radius R by $c=\frac{3R}{2G_N}$.

   Therefore, the injected energy density profile at time $t$ is given by
\be\label{TPtt}
T^{\mathcal{P}}_{tt}(x,t)  = T^{\mathcal{P}}(w) + \overline{T}^{\mathcal{P}}(\overline{w})
\ee
where $w=x+t, \; \overline{w}=x-t$. Moreover, assuming the LPO is performed at $t=0$, then the amount of the injected energy is
\be\label{master EA}
\Delta E_{\mathcal A}:= \int_{-\infty}^{\infty} dx\;  T^{\mathcal{P}}_{tt}(x,t=0) \;.
\ee
Note that $\Delta E_{\mathcal A}$ should be positive definite as required by the passivity.

   To have the explicit $T^{\mathcal{P}}$ and $\overline{T}^{\mathcal{P}}$ for the Banados geometry, we need to specify the LPO in CFT.  However, in CFT$_2$ the stress tensor for an excited state can be obtained by some conformal map from some geometry associated with the excited state to the (upper-half) complex plane (vacuum state)
\be
w \longrightarrow \xi:=\xi(w)\;.
\ee
so that
\be
 T^{\mathcal{P}} (w) = \frac{c}{12}\{\xi,w\}
\ee
where  $\{\xi, w\}:= \frac{\partial_w^3 \xi}{\partial_w \xi}-\frac{3}{2}(\frac{\partial_w^2\xi}{\partial_w\xi})^2$ is the Schwartzian derivative.  Therefore, one can construct the LPO which can be realized as some conformal map.

\begin{figure}
\includegraphics[width=.5\columnwidth]{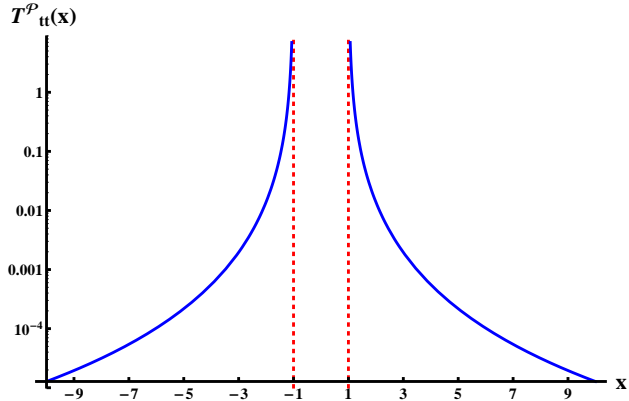}
\caption{ The energy density $T^{\mathcal{P}}_{tt} (x)$ with $u=-1$ and $v=1$. It extends everywhere with a power-law decay profile while away from the interval .}
\label{TPttfig}
\end{figure}

 In this work, we will focus on the one proposed in \cite{Numasawa:2016emc}, which is to project the local state on interval $\mathcal{A}:=\{x|u<x<v \}$ into state of no spatial entanglement. In \cite{Miyaji:2014mca,Miyaji:2015fia} it was shown that this kind of states are the Ishibashi states of boundary CFT.  Moreover, In \cite{Rajabpour:2015uqa,Rajabpour:2015xkj} this kind of LPOs is realized as a conformal map from UHP to a plane with a slit/cut, and has be further elaborated recently in \cite{Numasawa:2016emc} by also considering its holographic BCFT realization.  The slit/cut is the interval $\mathcal{A}$ on which the LPO acts. Explicitly, the conformal map is given by
\be
\xi(w)=\sqrt{ w-u \over v-w}
\ee
so that
\be\label{TP-zeroT}
 T^{\mathcal{P}} (w)={c \over 32} {(v-u)^2 \over   (w-u)^2(w-v)^2}\;.
\ee
The resultant energy density is plotted in Fig. \ref{TPttfig}. Note that the energy density is positive definite, and more importantly it extends everywhere with a power-law decay profile while away from the interval $\mathcal A$ despite that it is singular at the edge of the interval.

  In fact, the extensiveness of the positive injected energy density shown in  Fig. \ref{TPttfig} is a very peculiar feature from the QET point of view. This implies that the injected energy spread out immediately, possibly due to the long range entanglement structure of CFT's ground state. This is in analogue to the vacuum entanglement considered in \cite{Reznik:2002fz,Reznik:2003mnx,Reznik-Dirac} (see also \cite{Verdon-Akzam:2015tma}) for relativistic fields due to superoscillation phenomena \cite{superosci-0,superosci-1} \footnote{We thank M. Hotta for pointing this out to us.}. Furthermore, it also implies that there is a high chance for the distant Bob to gain energy extraction even without knowing the outcome of Alice's LPO. Later we will perform the holographic local unitary operation on the state dual to the Banados geometry with the stress tensor given by \eq{TP-zeroT}, and we will see this is indeed the case. %dg7
This procedure can be thought as applying a local unitary operation that
 excites the boundary gravitons and the cut-off surface, whose energy can then be extracted by another observer.     On the other hand, if the underlying theory is gapped, we may expect the injected energy density is short-ranged (exponentially decaying) so that the distant Bob may have rare chance to extract substantial energy out of the excited state caused by LPO.

  Note that the stress tensor profile \eq{TP-zeroT} is the same as the one for the insertion of twist operators of conformal dimension $\Delta={3\over 8}$ at $x=u$ and $x=v$. This means that this LPO can also be interpreted as the insertion of two twist operators at the edges of the interval. It then suggests that one may construct the other LPOs by inserting other primary operators at the edge of the interval. In fact, this is quite similar to the OPE blocks currently constructed in \cite{POVM-CFT}.  Moreover, by following the suggestive connection between LPO and twist operator insertion, we can generalize the construction of  \cite{Rajabpour:2015uqa,Rajabpour:2015xkj,Numasawa:2016emc} to the cases of multi-intervals. Then, by the usual trick of uniformization of multiple branch cuts on complex plane, we can obtain the stress tensor for these cases,
\be\label{master L}
 T^{\mathcal{P}} (w)={c\over 12} \sum_{i=1,2,\cdots,2N} \Big( {3 \over 8 (w-a_i)^2} + {p_i \over w-w_i} \Big)
\ee
where the accessory parameters $p_i$'s can be solved from the monodromy constraints  \cite{Faulkner:2013yia, Hartman:2013mia}.

  Finally, before we move on, we shall emphasize: it turns out that Bob's gain of extraction energy after performing the holographic LU to the injected energy density is universal, i.e., their ratio is independent of the form of LPO but depends only on the form of LU. Therefore, as long as the projected state is described by the Banados geometries, the universality holds for our holographic QET scheme.

 \subsection{Holographic local unitary operation}

The second step is to implement holographic LUs via surface/state duality \cite{Miyaji:2015yva,Miyaji:2015fia}. As sketched before,  we realize the holographic LUs by deforming the flat UV cutoff surface by a small local bump. We then evaluate the holographic stress tensor associated with the deformed UV surface, which is dual to a new CFT state with a LU acting on the old one associated with the flat UV cutoff surface.

\subsubsection{Holographic LU on ground state}

Before considering LU for holographic QET, we first consider the effect of LUs acting on the CFT ground state. That is, we will evaluate the holographic stress tensor in the pure AdS$_3$ space for a deformed UV surface which is different from the flat one by a local bump.

We start with the pure AdS$_3$ metric in Poincare coordinates
\be\label{hyperS}
ds^2:={R^2(dz^2 +dx^2-dt^2) \over z^2}\;.
\ee
To find the energy change due to the LUs in a region $\mathcal B$, it is just to evaluate the holographic stress tensor on a deformed UV cutoff surface with a local bump around $\mathcal B$, i.e., parametrized by
\be\label{deformedS}
r(z,x)=z-\phi(x)\;.
\ee
We will assume $\phi(x)$ is small.

The (regularized) holographic stress tensor on an UV hypersurface is given by \cite{Henningson:1998gx,Balasubramanian:1999re,deHaro:2000vlm}
\be
T_{\mu\nu}=\frac{1}{4 G_N}\left[ K_{\mu\nu} -K\gamma_{\mu\nu}-\frac{1}{R}\gamma_{\mu\nu}\right]
\ee
where $\gamma_{\mu\nu}$ and $K_{\mu\nu}$ are the pull-back metric and the extrinsic curvature for the hypersurface, and $K:=\gamma^{\mu\nu}K_{\mu\nu}$.  Given the hypersurface \eq{hyperS}, the result for the energy density is
\be\label{TLUtt}
T^{\mathcal{U}}_{tt}=\lim_{\epsilon \rightarrow 0} {c\over 6} \Big[ {1- (1+\phi'^2)^{-1/2} \over \epsilon^2}-{\phi'' (1+\phi'^2)^{-3/2} \over \epsilon}  \; \Big]
\ee
where $\epsilon$ is the UV cutoff and $'$ means taking derivative with respect to $x$.

 This result says that the energy density is divergent unless $\phi'=0$. It implies that if the hypersurface is not a minimal surface (a flat one in this case), the energy density blows up. On the other hand, if the hypersurface is a minimal one, the energy density is zero. The latter fact is consistent with the expectation of MERA where the state at each layer of the network should remain as the ground state at that energy scale. In Appendix, as a demonstration we discuss the pattern of the energy density profile due to inhomogeneous infinitesimal LU.

\subsubsection{Holographic LUs for QET}

Now we will consider the energy density profile of LUs derived from the bulk metric \eq{master metric} dual to the state after the particular LPOs specified by \eq{master L}. The procedure is the same as in the previous subsection except we now use the metric \eq{master metric} to evaluate the holographic stress tensor for the deformed UV cutoff surface \eq{deformedS}. Denote the energy density after LU acting on the dual state of the bulk metric \eq{master metric} by $T^{\mathcal{PU}}_{tt}$, then extraction energy density profile for QET is $\Delta \rho_{\mathcal B}:= T^{\mathcal{P}}_{tt}- T^{\mathcal{PU}}_{tt}$.

We choose to regularize $\Delta \rho_{\mathcal B}$ by treating the term $T^{\mathcal{U}}_{tt}$ as from the ``reference state" (without LPO) so that it should be subtracted away. %dg5
The regularization  is analogous to the prescription of Brown and York \cite{Brown:1992br} in the pre-AdS/CFT era for evaluating the quasilocal stress tensor in flat space by subtracting the divergent counter term of a reference boundary. In our case, this reference boundary is the surface dual to holographic LU, which yields a  divergent  stress tensor as a counter term.  In contrast, the stress tensor associated with the usual ``flat" boundary is finite.

\begin{figure}
\includegraphics[width=.52\columnwidth]{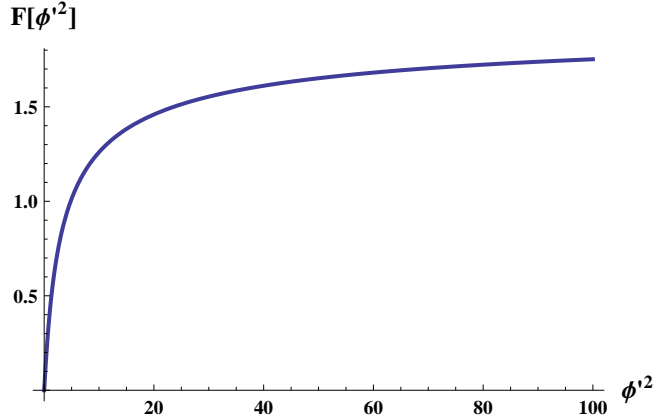}
\caption{ The functional $F\left[\phi'^2\right]$ is positive-definite and monotonically increasing. }
\label{F}
\end{figure}

 Follow the above procedure, we obtain the regularized extraction energy density profile for all orders of $\phi'^2$:
 \be
\label{regEd}
\Delta \rho^{(reg)}_{\mathcal B} = F\left[\phi'^2\right]T^{\mathcal{P}}_{tt}~,
\ee
where
\be\label{F-profile}
F\left[\phi'^2\right]:={4\Big(-1+\sqrt{1+\phi'^2}\Big)+ \phi'^2\Big(-5+4\sqrt{1+\phi'^2}\Big)\over 2\Big(1+ \phi'^2\Big)^{3/2} } ~.
\ee
It is straightforward to check that the profile of $F[\phi'^2]$ is positive definite, and the exact form is shown in Fig. \ref{F}. This implies that even for the finite LUs, the extraction energy density for holographic QET is positive definite, so is the extraction energy $\Delta E^{(reg)}_{\mathcal B} =\int_{-\infty}^{\infty} dx \; \Delta \rho^{(reg)}_{\mathcal B} $. One may expect this is related to some positive energy condition in dual gravity. In particular, our positive energy condition is a local one rather than the integrated ones as discussed in \cite{Flanagan:1997gn,Bousso:2015wca,Koeller:2015qmn}. Part of the reason is that the injected energy density $T^{\mathcal{P}}_{tt}$ by LPO is positive definite, which should be due to the feature of CFT and holographically is related to the positive energy condition of Banados geometry. This is in contrast to the non-CFT case, which usually yields oscillatory and non-positive definite results \cite{Hotta:2011xj}. On the other hand, the positivity of $F\left[\phi'^2\right]$ is not so transparent. It could be due to the fact that the local unitary operation is holographic dual to the graviton excitation, as previously commented, energy density of which should be positive definite as commonly expected.  However, the detailed picture relating to the positive energy theorem of dual gravity deserves further study.

Notice also that he form of $\Delta \rho^{(reg)}_{\mathcal B} $ does show an interesting and intuitive interplay between LPO and LU in the extraction energy density. It is a linear relation: the amount of extraction energy is proportional to the injected energy density by LPO, however, the efficiency depends on the profile of LU  at region $\mathcal B$. Moreover, the functional form of $F$ is universal, i.e., independent of the details of profile $\phi$.

It is interesting to observe the behaviors of $F\left[\phi'^2\right]$ in the small and large  $\phi'^2$ limit, i.e.,
\beq
\label{Fexp}
F[\phi'^2] \rightarrow \left\{\begin{array}{ll}
\frac{ 1}{2}\phi'^2,& \quad \mbox{for $\phi'^2\ll 1$} \\
2-\frac{5}{2|\phi'|}, & \quad \mbox{for $\phi'^2 \gg 1$}.
\end{array} \right.
\eeq
Note that $F$ saturates to the value $2$ for large enough $\phi'^2$. This implies that the efficiency of energy extraction is bounded, which is consistent with the second law of thermodynamics.

\section{Holographic QET at finite temperature}\label{sec 4}

In this section we generalize the above QET protocol to the finite temperature case. We will start with the planar BTZ black hole  \cite{Banados:1992wn}  with the following metric
\be
\label{BTZ}
ds^2=\frac{R^2}{z^2}\Big\{-(1-\frac{\pi^2 z^2}{\beta^2})^2dt^2+(1+\frac{\pi^2 z^2}{\beta^2})^2dx^2 + dz^2 \Big\}~,
\ee
where $\beta$ is the inverse temperature. Compared with the general form of Banados metric \eq{master metric} it tells $L(w)=\frac{\pi^2}{\beta^2}$, which yields the energy density of the dual CFT thermal state as
\be\label{TttBH}
T_{tt}^{\beta}=\frac{c \pi^2 }{3\beta^2}~.
\ee
To compute the change of energy density due to the application of LU on such holographic CFT's thermal state we need to evaluate the corresponding stress tensor denoted by $T_{tt}^{\mathcal{U}\beta}$
\be\label{TLUBH}
T_{tt}^{(reg) \mathcal{U} \beta}:=T_{tt}^{\mathcal{U}\beta}-T_{tt}^{\mathcal{U}}=(1-F[\phi'^2])T_{tt}^{\beta}.
\ee
the functional $F[\phi'^2]$ is given by \eq{F-profile} and $T_{tt}^{\mathcal{U}}$ is the divergent counter term at zero temperature given in \eq{TLUtt}. This substraction is sensible since the divergences are coming from the UV region where the surface does not feel the temperature.

\begin{figure}
\includegraphics[width=.52\columnwidth]{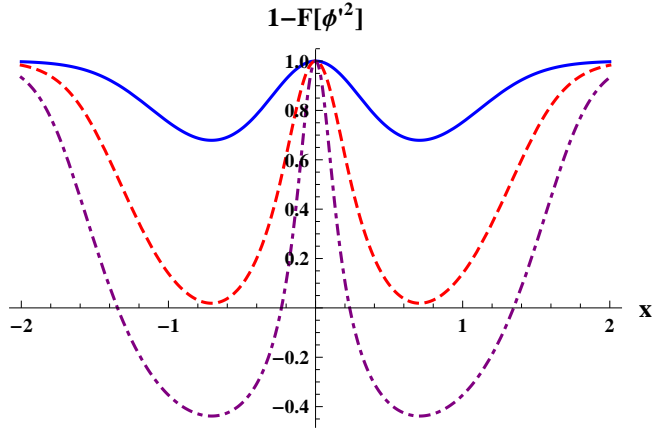}
\caption{ $T_{tt}^{(reg) \mathcal{U}\beta}(x) / T_{tt}^{\beta}:=1-F[\phi'^2(x)]$ for the Gaussian type \eq{Gaussian-h} with $\s=1$ and amplitude $A=1$ (blue-solid), $5/2$ (red-dashed) and $5$ (purple-dotdashed). Notice that for $A=5$ there are negative regions.}
\label{TUB}
\end{figure}

The change of the energy density $T_{tt}^{(reg) \mathcal{U}\beta}$ is positive definite for slowly varying surfaces $\phi'^2\ll 1$, since $F[\phi'^2] \simeq \frac{ 1}{2}\phi'^2$, and becomes negative for $\phi'^2 > 2(1+\sqrt{2})$. This implies that in the presence of the black hole horizon, that one can extract energy from the thermal state if the rate of the deformation is larger than some critical value, which is independent of the temperature. This is not unexpected since the surfaces are close to the UV region. In Fig. \ref{TUB} we show an example of the form of $T_{tt}^{(reg) \mathcal{U}\beta}(x)/T_{tt}^{\beta}$ for Gaussian $\phi$.

\subsection{Finite temperature LPO from a conformal map}

   Here we consider the specific LPO at finite temperature, which can be obtained by the following conformal map \cite{Numasawa:2016emc} \footnote{In the following we consider the Lorentzian version of the one given in \cite{Numasawa:2016emc} by changing their $\beta_H$ to $i \beta$. }
\be
\xi (w)=\sqrt{{\sin{{\pi \over \beta} (q-w)} \over \sin{{\pi \over \beta} (q+w)}}}\;.
\ee
Note that here we have set the Alice's interval to be ${\mathcal A}:=\{x|-q<x<q\}$.

\begin{figure}
\includegraphics[width=.48\columnwidth]{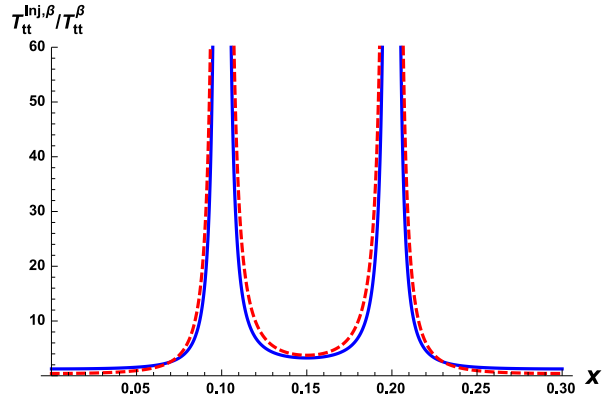}\hspace{.5cm}
\caption{The ratio of $T_{tt}^{\mathcal{P}\beta}(x)/T_{tt}^{\beta}(x)$ (blue-solid) and the  $T_{tt}^{Inj, \beta}(x)/T_{tt}^{\beta}(x)$ (red-dashed), in range of x smaller than the period $\beta$. Notice that both functions have the same poles.  Here $\beta=0.3$, $q=2$. }
\label{Fluc-TPBH}
\end{figure}

Using the Schwarzian derivative we can obtain the energy density due to LPO acting on $\mathcal{A}$ of the holographic CFT's thermal state, the result is
\be\label{TttPBH}
T_{tt}^{\mathcal{P}\beta}=\frac{11+\cos{\frac{4\pi q}{\beta}}-16\cos{\frac{2\pi q}{\beta}}\cos{\frac{2\pi x}{\beta}}+4\cos{\frac{4\pi x}{\beta}}}{8 \Big(\cos{\frac{2\pi q}{\beta}}-\cos{\frac{2\pi x}{\beta}}\Big)^2} \; T_{tt}^{\beta} \;.
\ee
The energy density is positive definite with a singularity at $|x|=q$ as in the zero temperature case, which can be seen from the denominator in the prefactor of \eq{TttPBH}.  We plot the periodic pattern in  Fig. \ref{Fluc-TPBH}.

Furthermore, we can subtract the $T_{tt}^{\mathcal{P}\beta}$ by the thermal state energy \eq{TttBH}, the result is the injected energy density due to LPO. We denote it as $T_{tt}^{Inj, \beta}$, which can be expressed as follows
\be\label{TinjBH}
T_{tt}^{Inj, \beta} = \frac{3 \sin^2{2\pi q \over \beta}}{4 \Big(\cos{\frac{2\pi q}{\beta}}-\cos{\frac{2\pi x}{\beta}}\Big)^2} \; T_{tt}^{\beta}
\ee
which is obviously positive semi-definite, and becomes zero for $q=0$ as expected. The typical pattern is also shown in Fig. \ref{Fluc-TPBH}, from which we find that $T_{tt}^{Inj, \beta}$ and $T_{tt}^{\mathcal{P}\beta}$ share similar patterns. Moreover, from \eq{TinjBH} it is easy to see that
\be
T_{tt}^{Inj, \beta}=0 \quad \mbox{so that} \quad T_{tt}^{\mathcal{P}\beta}=T_{tt}^{\beta}:=\frac{c \pi^2 }{3\beta^2}, \qquad \mbox{when} \quad  {q\over \beta}\in \mathbf{Z}\;.
\ee
Note that these particular patterns of zero injected energy happen only for $\beta \le q$, i.e., the relatively high temperature case.  This peculiar pattern could be holographically dual to that the geometrical operation associated with LPO is probing the near-horizon physics of the black hole as its size $q$ is comparable with the horizon scale $\beta$.

\subsection{Energy extraction at finite temperature}
Given the Banados geometry due to the LPO acting on the thermal state, the procedure for evaluating the extraction energy density for the holographic QET protocol is universal in our framework, so is the formal form the result.  To recapitulate, we first evaluate the total energy density after LPO and LU for the thermal state, and the result denoted by $T_{tt}^{\mathcal{P}\mathcal{U}\beta}$ is
\be
T_{tt}^{\mathcal{P}\mathcal{U}\beta}=T_{tt}^{\mathcal{U}}+(1-F\left[\phi'^2\right])T_{tt}^{\mathcal{P}\beta}.
\ee
We then evaluate the extraction energy density $\Delta \rho^{\beta}_{\mathcal B}:=T_{tt}^{\mathcal{P}\beta}-T_{tt}^{\mathcal{P}\mathcal{U}\beta}$  and obtain
\be
\Delta \rho^{\beta}_{\mathcal B}=-T_{tt}^{\mathcal{U}}+F\left[\phi'^2\right]T_{tt}^{\mathcal{P}\beta}.
\ee
Finally we subtract the divergent counter term $-T_{tt}^{\mathcal{U}}$ to obtain the regularized extraction energy density for QET, and the result is
\be
\Delta \rho^{(reg)\beta}_{\mathcal B}=F\left[\phi'^2\right] T_{tt}^{\mathcal{P}\beta}.
\ee
Note that it takes the same form as \eq{regEd} except by replacing $T_{tt}^{\mathcal{P}}$ with $T_{tt}^{\mathcal{P}\beta}$ as expected by the universality of our framework.  Obviously,  the energy extraction is positive definite as guaranteed by the positive-definiteness of both $F\left[\phi'^2\right]$ and $T_{tt}^{\mathcal{P}\beta}$.

\section{Conclusions}

In this paper we propose a tentative holographic scheme for the QET protocol. In the spirit of holography we try to realize all the quantum operations involved in the protocol as geometrical as possible. As we have no concrete way at this moment to realize POVM in CFTs or the holography, we choose to modify the conventional QET protocol by considering extraction of energy out of a particular excited state obtained by local projection. In this way we do not need the correlated the measurement outcome with Bob's local operation when extracting energy. Despite that, our protocol still succeeds in gaining the energy extraction. The success of our simplifying QET protocol can be attributed to the following facts: (a) the injected energy density due to local projection is positive semi-definite; and (b) the proportionality between Bob's extraction energy density and the energy density of the excited state after Alice's LPO is also positive definite. Moreover, this proportionality is universal in the sense that it does not depend on Alice's local projection.  It is not clear if such positive semi-definiteness is universal or not. If so, it may be related to some sort of the positive energy condition. In our work we have shown that the positive semi-definiteness holds even with the presence of bulk black hole. It is very interesting to see the possible connection between the quantum information inequality and the positive energy condition.

 Another interesting perspective of our study is the demonstration how the surface/state duality (or AdS/MERA duality) can help to realize some quantum information task. In our case, we argue that the deformation of the UV surface can be thought as the local (unitary) operations in accordance with the surface/state duality. It would be interesting to understand more precisely what is the dual operation in the CFT side for these surface deformation. This reinforces the emerging point of view of bulk gravity from the quantum circuit based on the underlying CFT. Besides, the consideration of deforming UV surface also rises the new technical issue of finding a more covariant way of regularizing the stress tensor associated with such kind of surface. For convenience we adopt the old regularization method as Brown and York \cite{Brown:1992br}. %dg5, and still can obtain the sensible answer.
 We believe that our efforts in this work, in connection of quantum information tasks to holography are just part of the first step along this direction and more studies are needed.

\section*{Acknowledgement}
FLL would like to thank M. Hotta and T. Takayanagi for helpful discussions and their hospitality during his visit of Tohuku U. and YITP. We also thank B. Czech for helpful discussions and collaborations on the related topics. This work is supported by Taiwan Ministry of Science and Technology through Grant No.~103-2112-M-003 -001 -MY3  and 103-2811-M-003 -024 and is
supported in part by  the National Center of Theoretical Science
(NCTS) and the grants  101-2112-M-007-021-MY3 and 104-2112-M-007 -001 -MY3 of the Ministry of Science and
Technology of Taiwan.

\bigskip

\section*{Appendix: Energy density profile for infinitesimal holographic local operations}

   Due to the divergent expression \eq{TLUtt} for the energy density profile caused by the
inhomogeneous infinitesimal LU, it is hard to find out the typical pattern of this energy density profile. However, we can consider the infinitesimal LU which has the amplitude of UV cutoff to see the pattern. For this purpose, we can expand the expression of  \eq{TLUtt} up to $\mathcal{O}(\phi^2)$, and the result  is
\be
T^{\mathcal{U}}_{tt}=\lim_{\epsilon \rightarrow 0} {c\over 12} \Big[ \; \Big( {\phi' \over \epsilon} \Big)^2 - 2 { \phi'' \over \epsilon} \; \Big]\;.
\ee
Furthermore, we require the overall size of $\phi$ is $\mathcal{O}(\epsilon)$, i.e., $\phi(x)=\epsilon h(x)$, and then
\be
T^{\mathcal{U}}_{tt}=  {c\over 12}  \Big( h'^2 - 2  h''  \; \Big)\;.
\ee
This can be understood as the energy density profile induced by LUs on a region with few lattice sites if we discretize the space where CFT lives with lattice spacing of $\mathcal{O}(\epsilon)$.

\begin{figure}
\includegraphics[width=.52\columnwidth]{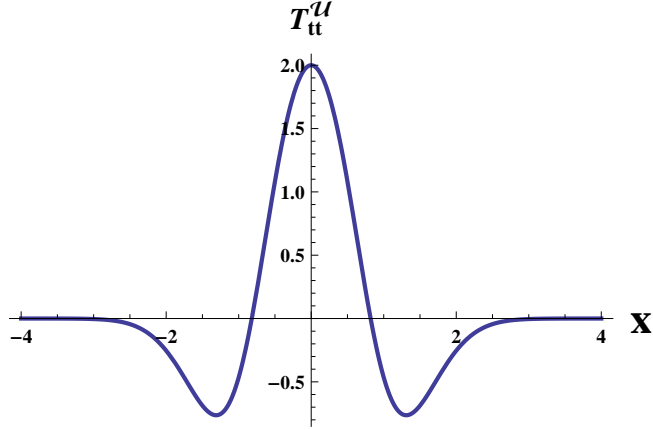}
\caption{The energy density profile $T^{\mathcal{U}}_{tt}(x)$ for $\phi(x)=\epsilon \; e^{-x^2}$. }
\label{Ttt-LU-n}
\end{figure}

   If we consider the bump to be in the Gaussian form
\be\label{Gaussian-h}
h(x)=\exp \{\frac{-x^2}{\sigma^2} \}\;,
\ee
then the typical energy density is shown in Fig. \ref{Ttt-LU-n}, where we see some region of negative energy density, but overall it is positive. This means that the Gaussian LU injects energy. This is in contrast to the energy density of LPO, e.g., \eq{master L}, which is extensive and positive definite, but with vanishing regularized energy.

\end{document}